\documentclass{llncs}

\usepackage{latexsym}
\usepackage{epsfig}
\usepackage{amssymb}
\usepackage{array}
\usepackage{float}

\graphicspath{{figures/}{./}}

\floatstyle{ruled}
\newfloat{algo}{thp}{lop}
\floatname{algo}{Algorithm}

\newcommand{\IF}[1]{\textbf{if} {#1}}
\newcommand{\THEN}{\textbf{then} }
\newcommand{\ELSE}{\textbf{else} }

\newcommand{\ENDIF}{\textbf{endif} }

\newcounter{ALGOline}
\newcounter{ALGOAlgo}
\newcommand{\LINESEP}{.}
\newcommand{\LINESTYLE}{\tiny}
\newcommand{\INITALGO}[1]{\setcounter{ALGOAlgo}{#1}}
\newcommand{\INITLINE}{\setcounter{ALGOline}{1}}

\newcommand{\AL}{
    \LINESTYLE{\arabic{ALGOAlgo}\LINESEP
    \ifnum \value{ALGOline}<10 0\fi
    \LINESTYLE{\arabic{ALGOline}}
    \addtocounter{ALGOline}{1}
}}

\newcommand{\LA}{<\!{\tt Leader}\! \leadsto\! {\tt Agreement}\!>} 
\newcommand{\AT}{<\!{\tt Agreement}\! \leadsto\! {\tt Pattern}\!>}

\newcommand{\BEGLIST}{\begin{list}{}{\partopsep -3pt \parsep -2pt \listparindent -0pt \labelwidth .5in}}
\newcommand{\ENDLIST}{\end{list}}

\newcommand{\ie}{\emph{i.e., }}

\begin{document}

\INITALGO{0}
\INITLINE

\title{Leader Election Problem Versus Pattern Formation Problem} 

\titlerunning{Leader Election Problem Versus Pattern Formation Problem}

\author{Yoann Dieudonn\'e\inst{1} \qquad Franck Petit\inst{2} \qquad Vincent Villain\inst{1}}

\authorrunning{Dieudonn\'e and Petit and Villain}   

\tocauthor{Yoann Dieudonn\'e and Franck Petit and Vincent Villain}

\institute{MIS,  Universit\'{e} de Picardie Jules Verne Amiens, France
\and
LIP6/Regal, Universit\'e Pierre et Marie Curie, INRIA, CNRS, France}
\date{}
\maketitle


\begin{abstract}
Leader election and arbitrary pattern formation are fundamental tasks for a set of 
autonomous mobile robots. 
The former consists in distinguishing a unique robot, called the leader.
The latter aims in arranging the robots in the plane to form any given pattern. 
The solvability of both these tasks turns out to be necessary in order to achieve more complex tasks. 

In this paper, we study the relationship between these two tasks in a model, called $CORDA$, 
wherein the robots are weak in several aspects. In particular, they are \emph{fully asynchronous} and they have no direct means of communication.
They cannot remember any previous observation nor computation performed in any previous step.  
Such robots are said to be \emph{oblivious}. 
The robots are also \emph{uniform} and \emph{anonymous}, i.e, they all 
have the same program using no global parameter (such as an identity) allowing to differentiate 
any of them. Moreover, we assume that none of them share any kind of common coordinate mechanism or common 
sense of direction and we discuss the influence of a common \emph{handedness} (i.e., \emph{chirality}).

In such a system, Flochini \emph{et al.} proved in \cite{FPSW08} that it is possible to elect a leader for $n\geq3$ robots if it is possible to form any pattern for $n\geq3$. In this paper, we show that
the converse is true for $n\geq4$ when the robots share a common handedness and for $n\geq5$ when they do not. Thus, we deduce that with chirality (resp. without chirality) both problems are equivalent for $n\geq4$ (resp. $n\geq5$) in $CORDA$.

\end{abstract}

\section{Introduction}
Mobile robots working together to perform cooperative tasks in a given
environment is an important, open area of research.
{\em Teams} (or, {\em swarms}) of {\em mobile robots} provide the ability to measure properties, 
collect information and act in a given physical environment.
Numerous potential applications exist for such multi-robot systems, 
to name only a very few: environmental monitoring, large-scale construction,
risky area surrounding or surveillance, and exploration of awkward environments.

In a given environment, the ability for the swarm of robots to succeed in 
the accomplishment of the assigned task greatly depends on ($1$) {\em global} properties
assigned to the swarm, and ($2$) {\em individual} capabilities each robot has.  
Examples of such global properties are the ability to distinguish among themselves at least one (or, more) 
robots ({\em leader}), to agree on a common global direction ({\em sense of direction}), 
or to agree on a common handedness ({\em chirality}).  The individal capacities of a robot are its 
moving capacities and its sensory organs.

To deal with cost, flexibility, resilience to dysfunction, and autonomy,
many problems arise for handling the distributed coordination of 
swarms of robots in a {\em deterministic} manner. 
This issue was first studied in~\cite{SY96,SY99}, mainly motivated by the minimal level of ability 
the robots are required to have in the accomplishment of basic cooperative tasks.
In other words, the feasibility of some given tasks is addressed assuming swarm of autonomous robots 
either devoid or not of capabilities like (observable) identifiers, direct means of communication, 
means of storing previous observations, sense of direction, chirality, etc. 
So far, except the ``classical'' \emph{Leader Election Problem}~\cite{CG07,DP07,FPSW08,P02}, 
most of the studied tasks are geometric problems, so that \emph{Arbitrary Pattern Formation Problem}, \emph{Line Formation}, 
\emph{Gathering}, and \emph{Circle Formation}---refer 
to~\cite{CP08,DK02,DLP08,FPSW00,FPSW08,K05,SY99} for these problems.

In this paper, we concentrate on two of the aforementioned problems: Leader Election Problem ($LEP$) and Arbitrary Pattern Formation Problem ($APFP$).

\begin{definition}[$LEP$]\cite{FPSW08}
\label{def:lep}
Given the positions of $n$ robots in the plane, the $n$ robots are able to deterministically agree on the same robot $L$ called the leader. Initially, the robots are in arbitrary positions, with the only requirement that no two robots are in the same position.
\end{definition}

\begin{definition}[$APFP$]\cite{FPSW08}
\label{def:pfp}
The robots have in input the same pattern, called the target pattern $\mathcal{P}$, described as a set of positions in the plane given in lexicographic order (each robot sees the same pattern according to the direction and orientation of its local coordinate system). They are required to form the pattern: at the end of the computation, the positions of the robots coincide, in everybody's local view, with the positions of $\mathcal{P}$, where $\mathcal{P}$ may be translated, rotated, and scaled in each local coordinate system. Initially, the robots are in arbitrary positions, with the only requirement that no two robots are in the same position, and that the number of positions prescribed in the pattern and the number of robots are the same.
\end{definition}

The issue of whether $APFP$ or $LEP$ can be solved or not
according to some capabilities of the robots is addressed in~\cite{FPSW08}. Not surprisingly, both problems are not deterministically solvable in general, due to the anonymity and the disorientation of the robots. This is especially true for $LEP$ for which the impossibility of breaking a possible symmetry makes $LEP$ unsolvable. For that matter, in \cite{DP07} the authors provide a complete characterization (necessary and sufficient conditions) on the robots positions to elect a leader in a deterministic way.

A first relationship between $APFP$ and $LEP$ is given by the following theorem:

\begin{theorem}\cite{FPSW08}
\label{theo:implique}
If it is possible to solve $APFP$ for $n\geq3$ robots, then $LEP$ is solvable too.
\end{theorem}

Naturally, an interesting question arises from the above theorem: ``{\em Is the converse true?}''. In other terms:  ``{\em With robots devoid of sense of direction, 
does $APFP$ becomes solvable whenever the robots have the possibility to 
distinguish a unique leader?}''
In \cite{YS10}, the authors provide a positive answer to this question assuming that robots have a common handedness.
The latter allows to infer the orientation of the $x$-axis once the orientation of the $y$-axis is given. Their result holds in the semi-synchronous model (SSM), {\it a.k.a.} Model SYm in
the literature.

In this paper, we show that this result also holds for $n\geq4$ (resp. $n\geq5$) robots, in a fully asynchronous model called CORDA, if the robots have a common handedness (resp. they do not).
Combined with Theorem~\ref{theo:implique}, we deduce that Leader
Election and Pattern Formation are two {\em equivalent} problems in $CORDA$ for $n\geq4$ robots with a common handedness (for $n\geq5$ without it), in the precise sense
that, the former problem is solvable if and only if the latter problem is solvable.

The rest of the paper is organized as follows: In Section~\ref{sec:model}, we describe the distributed
systems.  The proof of equivalence with chirality is given 
in Section~\ref{sec:algo} for any $n\geq 4$ by providing an algorithm working in CORDA. The case without chirality is discussed in Section~\ref{sec:sans}.
Finally, we make concluding remarks in Section~\ref{sec:concl}.

A preliminary version of this paper appears in \cite{DieudonnePV10}.

\section{Model}
\label{sec:model}

We adopt the model CORDA (COordination and control of a set of Robots in a totally Distributed and Asynchronous environment) introduced in \cite{FPSW99}.
The \emph{distributed system} considered in this paper consists of $n$ robots  
$r_{1}, r_{2},\cdots , r_{n}$---the subscripts $1,\ldots ,n$ are used for notational purpose only.
Each robot $r_{i}$ is viewed as a point in a two-dimensional 
space unbounded and devoid of any landmark.  When no ambiguity arises, $r_{i}$ also denotes the 
position in the plane occupied by that robot. Each robot has its own local coordinate system and unit measure.  
The robots do not agree on the orientation of the axes of their local coordinate system, 
nor on the unit measure.

\begin{definition}[Sense of Direction]
\label{def:sod}
A set of $n$ robots has sense of direction if the $n$ robots agree on a common
direction of one axis ($x$ or $y$) and its orientation.  
The sense of direction is said to be \emph{partial}
if the agreement relates to the direction only ---ie.  
they are not required to agree on the orientation. 
\end{definition}

In the rest of this paper, we assume that the robots have no sense of direction and we discuss the influence of chirality.

%
Given an $x$-$y$ Cartesian coordinate system, the \emph{handedness} is the way in which 
the orientation of the $y$ axis (respectively, the $x$ axis) is inferred according to 
the orientation of the $x$ axis (resp., the $y$ axis).  

\begin{definition}[Chirality]
A set of $n$ robots has chirality if the $n$ robots share the same handedness.
\end{definition}

The robot's life is viewed as an infinite sequence of cycles. Each cycle is a sequence of four states Wait-Observe-Compute-Move characterized as follows.
 
\paragraph{Life cycle.} Initially, a robot is in the waiting state ({\bf Wait}). Asynchronously and
independently from other robots, it observes its surroundings ({\bf Observe}) by using its sensors. The sensors return a set of all the positions occupied by at least one robot, with respect to its own
coordinate system. Then, from its new observations the robot computes its next location ({\bf Compute}) according to a given protocol which is the same one for all the robots. Once the computation is done, the robot moves towards its new location ({\bf Move}). The distance traveled by a robot in a cycle is unpredictable and thus, the robot may stop its motion before reaching the computed location. However, the distance traveled by a robot r in a move is neither infinite nor infinitesimally small. In particular, there exists a constant $\sigma_r >0 $ such that if the destination point is closer than $\sigma_r$, r will reach it; Otherwise, r will move towards it of at least $\sigma_r$. Finally, the robot returns to the waiting state. It is assumed that the amount of time spent in each phase of a cycle is finite but unpredictable and may be different for each cycle and for each robot. That is why the robots are considered to be fully asynchronous. 

Finally we assume that the robots are \emph{uniform} and \emph{anonymous}, i.e, they all have the same program using no 
local parameter (such as an identity) 
allowing to differentiate any of them. Moreover, they have no direct means of communication and they are \emph{oblivious}, i.e.,
none of them can remember any previous observation nor computation performed 
in any previous cycles.

\section{Equivalence for $n\geq4$ with chirality}
\label{sec:algo}

In this section we prove the main result of this paper:

\begin{theorem}
\label{theo:result4}
In $CORDA$, assuming a group of $n \geq 4$ robots having chirality and devoid of 
any kind of sense of direction, $LEP$ is solvable if and only if $APFP$ is solvable.
\end{theorem}

To prove Theorem~\ref{theo:result4}, from Theorem~\ref{theo:implique}, it remains to show the following lemma:

\begin{lemma}
\label{lem:result4}
In $CORDA$, assuming a group of $n \geq 4$ robots having chirality and devoid of 
any kind of sense of direction, if $LEP$ is solvable, then $APFP$ is solvable.
\end{lemma}

The remainding of this section is devoted to prove Lemma~\ref{lem:result4} by providing a protocol that forms an arbitrary target pattern assuming that, initially the robots are in a \emph{leader configuration}, wherein the robots are able to deterministically elect a leader.

The overall idea of our algorithm consists of the three following main steps: 
First, by moving to some appropriate positions, the robots build a kind of global coordinate system.  
Next, they compute the final positions to occupy in order to form the pattern. 
Finally, the robots carefully move towards these final positions, while maintaining the global coordinate system invariant.
In the next subsection (Subsection~\ref{toto}), we provide basic definitions and properties leading to describe what is an (equivalent) agreement configuration. Then, in Subsection~\ref{titi}, we will give the distributed algorithm with its correctness proof.

\subsection{Agreement Configuration}
\label{toto}

In the rest of this paper, we assume the set of all the positions $\mathcal{Q}$ occupied by the robots in the plane is the set of all the coordinates expressed in a cartesian coordinate system $\mathcal{S}$ which is unknown for all the robots. However, all the coordinates $\mathcal{Q}$ expressed in $\mathcal{S}$ coincide with all the cordinates $\mathcal{Q}$ expressed in everybody's local system where $\mathcal{Q}$ may be translated, rotated or scaled.

\begin{definition}[Smallest enclosing circle]\cite{DK02} 
\label{def:sec}
Given a set $\mathcal{Q}$ of $n\geq2$ positions $p_1,p_2,\cdots,p_n$ on the plane, the smallest enclosing circle of $\mathcal{Q}$ , called $SEC(\mathcal{Q})$, is the smallest circle enclosing all the positions in $\mathcal{Q}$.
\end{definition}

When no ambiguity arises, $SEC(\mathcal{Q})$ is shortly denoted by $SEC$ and \\ $SEC(\mathcal{Q})\cap\mathcal{Q}$
indicates the set of all the positions both on $SEC(\mathcal{Q})$ and $\mathcal{Q}$. Besides, we say that a robot $r$ is inside $SEC$ if and only if $r$ is not located on the circumference of $SEC$. In any configuration $\mathcal{Q}$, $SEC$ is unique and can be computed in linear time~\cite{M83}.
%
Note that since the robots have the ability of chirality, they are able to agree on a common orientation of $SEC$, denoted $\circlearrowright$, in the sequel referred to as the clockwise direction.


\begin{property}\cite{W91}
\label{pro:sec}
$SEC$ passes either through two of the positions that are on the same diameter (opposite positions), or through at least 
three positions. $SEC$ does not change by eliminating or adding positions that are inside it. $SEC$ does not change by adding positions on its boundary. However, it may be possible that $SEC$ changes 
by either eliminating or moving positions on its circumference.
\end{property}

\begin{figure}[!htbp]
  \begin{center}
    \begin{minipage}[[t]{.3\linewidth}
      \centering
      \epsfig{file=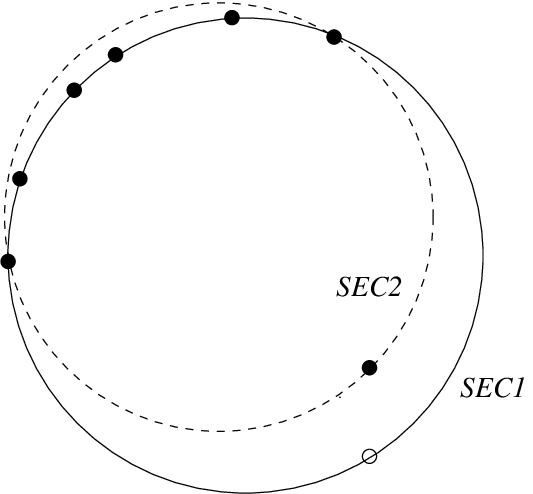, width=0.5\linewidth}\\
      {\scriptsize ($a$) Critical (white) robot cannot be deleted without changing $SEC$.}
    \end{minipage}
    \begin{minipage}[[t]{.6\linewidth}
      \centering
      \epsfig{file=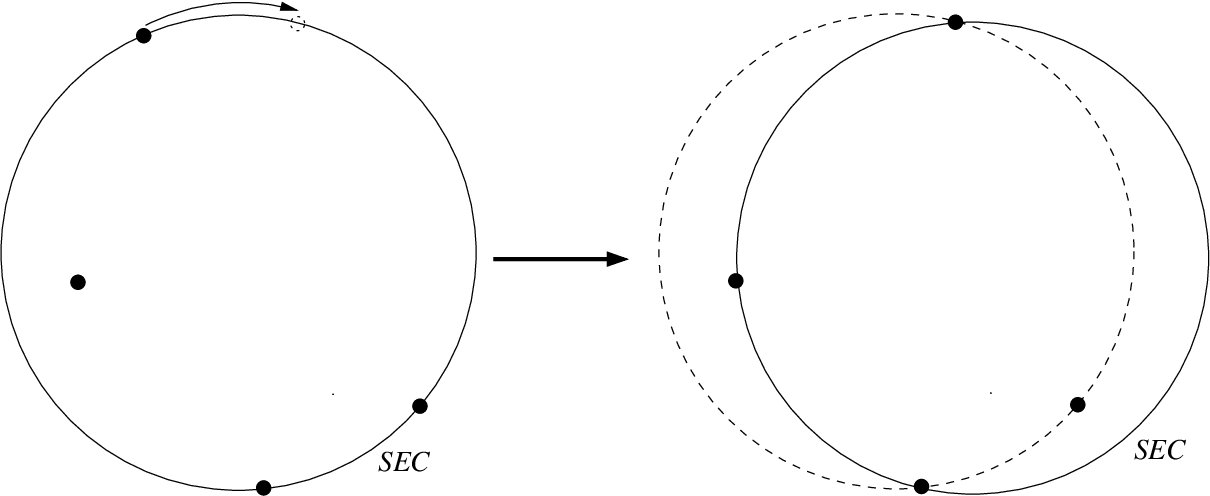, width=0.6\linewidth}\\
      {\scriptsize ($b$) An example showing how $SEC$ may change by moving one robot.}
     \end{minipage}
  \end{center}
  \caption{Examples illustrating Property \ref{pro:sec}}.
\label{figure123}
\end{figure}

Examples showing the latter assertion of Property~\ref{pro:sec} are proposed in Figure~\ref{figure123}.

\begin{definition}[Critical position]\cite{K05}
\label{def:critical}
Given a set $\mathcal{Q}$ of distinct positions. We say that a position $p$ is critical iff $SEC(\mathcal{Q})\ne SEC(\mathcal{Q}\setminus\{p\})$.
\end{definition}

An example of such a critical robot is given by Figure~\ref{figure123}, Case~($a$). 
According to Property~\ref{pro:sec}, a critical position cannot be inside $SEC$. So, we have the following corollary:

\begin{corollary}
\label{cor:cri}
Let $\mathcal{Q}$ be a configuration. If there exists a critical position $p$ in $\mathcal{Q}$, then $p$ is on the circumference of $SEC(\mathcal{Q})$.
\end{corollary}

Before giving other properties about critical positions, we need to define extra notions.

\begin{definition}[$adjacent(r,C,\circlearrowright)$]
Given a circle $C$ and a group of robots located on it, we say that $r'=adjacent(r,C,\circlearrowright)$ if $r'$ is the next robot on $C$ just after $r$ in the clockwise direction.
\end{definition}

In the same way, we can define $adjacent(r,\circlearrowleft)$ in the counterclockwise direction. When no ambiguity
arises, $adjacent(r,C,\circlearrowright)$ is shortly denoted by \\ $adjacent(r,\circlearrowright)$. Sometimes, if $r'=adjacent(r,\circlearrowright)$, we simply say that $r'$ and $r$ are \emph{adjacent}.

\begin{definition}[$angle(p,c,p',\circlearrowright)$]
Given a circle $C$ centered at $c$ and two points $p$ and $p'$ located on it,
$angle(p,c,p',\circlearrowright)$ is the angle centered at $c$ from $p$ to $p'$ in the clockwise direction.
\end{definition}

In the same way, we can define $angle(p,c,p',\circlearrowleft)$ in the counterclockwise direction.

The following properties are fundamental results about smallest enclosing circles:

\begin{lemma}\cite{CP02}
\label{lemma:sec}
Let $r_i,r_j$ and $r_k$ be three consecutive robots on $SEC$ centered at $c$ such that $r_j=adjacent(r_i,\circlearrowright)$  and $r_k=adjacent(r_j,\circlearrowright)$. 
If $angle(r_i,c,r_k,\circlearrowright)\leq180^o$, 
then $r_j$ is non-critical and $SEC$ does not change by eliminating $r_j$.
\end{lemma}

\begin{corollary}
\label{theorem1}
Let $SEC(\mathcal{Q})$ be the smallest circle enclosing all the positions in $\mathcal{Q}$. For all couple of positions $r_i$ and $r_j$ in $SEC(\mathcal{Q})\cap\mathcal{Q}$ such that $r_j=adjacent(r_i,\circlearrowright)$, we have $angle(r_i,c,r_j,\circlearrowright)\leq180^o$. 
\end{corollary}

\begin{lemma}\cite{CP02}
\label{lem:critical}
Given a smallest enclosing circle with at least four robots on it, there exists at least one robot which is not critical.
\end{lemma}

\begin{definition}[Concentric Enclosing Circle]
\label{def:conc}
Given a set $P$ of distinct positions. We say that $C^P$ is a concentric enclosing circle if and only if it is centered at 
the center $c$ of $SEC$, has a radius strictly greater than zero and it passes through at least one position in $P$.
\end{definition}

In the following, $\mathcal{SC}^P$ and  $|\mathcal{SC}^P|$ respectively denote the set of all the concentric enclosing circle in $P$ and its cardinality. For some $k$ such that $1\leq k \leq |\mathcal{SC}^P|$, $C^P_k$  indicates the $k^{th}$ greatest concentric enclosing circle in $P$ and $\bigcup_{i=1}^{k} C^{P}_i$ is the set of the $k$ first greatest enclosing circles in $P$. Moreover, we assume that a position (or robot) located inside a concentric enclosing circle $C^P_k$ is not on the circumference of $C^P_k$. $C^{P}_i\cap P$ indicate the set of all the positions both on $C^{P}_i$ and $P$. 

\begin{remark}
 From Definition~\ref{def:conc}, $SEC$ is the greatest concentric enclosing circle of $SC$ (i.e., $SEC=C_1$) and the center of $SEC$ cannot be a concentric enclosing circle.
\end{remark}

 From Definition~\ref{def:conc}, we can introduce the notion of \emph{agreement configuration}:
\begin{definition}[Agreement Configuration]
\label{def:agree}
A configuration $\mathcal{Q}$ is an agreement configuration if and only if both following conditions hold: 
\newline \noindent 1. There exists a robot $r_l$ in $\mathcal{Q}$ such that $r_l$  is the  unique robot located on the smallest concentric enclosing circle $C^{\mathcal{Q}}_{|\mathcal{SC}^\mathcal{Q}|}$,
\newline \noindent 2. There is no robot at the center of $SEC(\mathcal{Q})$.
\end{definition}

In an agreement configuration, $r_l$ is called the $\emph{leader}$.

\begin{definition}
\label{def:equiv}
Two agreement configuration $\mathcal{Q}_1$ and $\mathcal{Q}_2$ is said to be \emph{equivalent} if and only if both following conditions hold:
\newline \noindent 1.  $SEC(\mathcal{Q}_1)$ and $SEC(\mathcal{Q}_2)$ are superimposed.
\newline \noindent 2. Let $c_1$ and $c_2$ be respectively the center of $SEC(\mathcal{Q}_1)$ and the center of \\
 $SEC(\mathcal{Q}_2)$. Let $r_{l1}$ and $r_{l2}$ be respectively the leader in $\mathcal{Q}_1$ and the leader in $\mathcal{Q}_2$. $[c_1,r_{l1})$ and $[c_2,r_{l2})$ are superimposed.
\end{definition}


\subsection{The protocol}
\label{titi}

Starting from a leader configuration, the protocol, shown in Algorithm~\ref{algo:main}, allows to form any target pattern $\mathcal{P}$. It is a compound of two procedures presented below:
\newline \noindent 1. Protocol~$\LA$ transforms an arbitrary leader configuration into an agreement configuration;
\newline \noindent 2. Protocol~$\AT$ transforms an agreement configuration into a pattern $\mathcal{P}$.

\begin{algo}[htb!]
\begin{footnotesize}
\begin{tabbing}
  xxxxx \= xxxxx \= xxxxx \= xxxxx \= xxxxx \= xxxxx \= xxxxx \= xxxxx \= xxxxx \= \kill 
$\mathcal{P}:=$ the target pattern ;\\
\IF{the robots do not form the target pattern}\\
\THEN \> \IF{the robots do not form an agreement configuration}\\ 
      \> \THEN  Execute $\LA$;\\
      \> \ELSE  Execute $\AT$;
\end{tabbing}
\end{footnotesize}
\caption{
          Form an arbitrary pattern starting from a leader configuration ($n\geq4$). 
          \label{algo:main}}
\end{algo}



\paragraph{Procedure~$\LA$.}

\begin{algo}[htb!]
\begin{footnotesize}
\begin{tabbing}
  xxxxx \= xxxxx \= xxxxx \= xxxxx \= xxxxx \= xxxxx \= xxxxx \= xxxxx \= xxxxx \= \kill 
$\mathcal{Q}:=$ the configuration where the robots currently lies; \\
$r_l:= Leader(\mathcal{Q})$; \\
$c:=$ center of $SEC(\mathcal{Q})$\\
\IF {$r_l$ is located at $c$} \\
\THEN \> $r_k:=$ the closest robot to $c$ $\in \mathcal{Q}\setminus\{r_l\}$;\\ 
        \> $p:=$ the middle of the segment $[r_l;r_k]$;\\
	\> \IF {I am $r_l$} \THEN $MoveTo(p,\rightarrow)$;\\
\ELSE \> \IF {$r_l$ is not critical} \\
      \> \THEN \> $p:=$ the middle of the segment $[r_l;c]$;  \\
	\> \> \IF {I am $r_l$} \THEN $MoveTo(p,\rightarrow)$;\\
\> \ELSE \> /* $r_l$ is critical and $r_l$ is on $SEC$*/\\
 \>     \> $r_k:=$ the first non-critical robot starting from $r_l$ on $SEC$ in clockwise.\\   	
 \>     \> \IF{I am $r_k$}\\
  \>    \> \THEN \>  $p:=$ the middle of the segment $[r_k;c]$; $MoveTo(p,\rightarrow)$; \ENDIF
\end{tabbing}
\end{footnotesize}
\caption{Procedure~$\LA$ for any robot $r_i$ in an arbitrary leader configuration \label{algo:LA}}
\end{algo}

In a leader configuration, we have the following corollary:

\begin{corollary}\cite{DP07}
\label{corollaire}
If the robots are in a leader configuration, then they can distinguish a unique leader which is one of the closest robot to the center of the smallest enclosing circle of the configuration, provided that they share the property of chirality.
\end{corollary}

So, from Corollary~\ref{corollaire}, we know that we can distinguish a unique robot $r_l$, called the leader, which is 
one of the robots closest to the center $c$ of $SEC(\mathcal{Q})$. However, according to Definition~\ref{def:agree},
if $r_l$ is at the center of $SEC(\mathcal{Q})$ or if $r_l$ is not the unique robot closest to the center of
$SEC(\mathcal{Q})$, $\mathcal{Q}$ is not an agreement configuration. In that case, Procedure~$\LA$ allows to transform the leader configuration into an agreement configuration. Algorithm~\ref{algo:LA} describes Procedure~$\LA$.
In Algorithm~\ref{algo:LA}, we use two subsoutines: $Leader(\mathcal{Q})$ and $MoveTo(p,\rightarrow)$. The former returns the unique leader from a leader configuration $\mathcal{Q}$. The latter allows a robot $r$ to move towards the point $p$, using a straight movement.

\subsubsection{Procedure~$\AT$}

Intuitively, once the robots are in an agreement configuration, they can also agree on their final positions---refer to Definition~\ref{def:map}. Then, some selected robots (Definition~\ref{def:extra}) begin to occupy
them, starting from those situated on $SEC$, and then on all the circles concentric to $SEC$ from the largest to the
smallest passing through at least one of the final positions---refer to Definition~\ref{ksub}. During this phase, the final positions are maintained unchanged, by
making sure that the robots remain in an equivalent agreement configuration until the pattern is formed. In particular, we make sure that no angle above $180^o$ is created on $SEC$---otherwise, according to Corollary~\ref{theorem1} $SEC$
changes---and that the leader of the agreement configuration remains the unique closest robot from the center of
$SEC$ and do not leave the radius where it is located.

Before presenting Procedure~$\AT$ shown in Algorithm~\ref{algo:AT}, we need the following definitions:

\begin{definition}[$Map(\mathcal{Q},\mathcal{P})$]
\label{def:map}
Let $\mathcal{Q}$ and $\mathcal{P}$ be respectively an agreement configuration formed by the robots in the plane and a target pattern. \\ $Map(\mathcal{Q},\mathcal{P})$ is the set of all the final positions $\mathcal{P}$ expressed in the plane where the robots currently lie and computed as follows:
\newline \noindent 1. First, the center of $SEC(\mathcal{P}))$ is translated to the center of $SEC(\mathcal{Q})$. 
\newline \noindent 2. Then, let $o,c,r_l$ and $s$ be respectively the center of $SEC(\mathcal{Q})$, the center of $SEC(\mathcal{P})$, the leader in $\mathcal{Q}$ and the first non-critical position (in the lexicographic order) located on the smallest concentric enclosing circle of $\mathcal{P}$. $\mathcal{P}$ is rotated so that the half-line $[o,r_l)$ is viewed as the half-line $[c,s)$.
\newline \noindent 3. Finally, $\mathcal{P}$ is scaled with respect to the radius of $SEC(\mathcal{Q})$ in order that all the distances are expressed according to the radius of $SEC(\mathcal{Q})$. In particular $SEC(\mathcal{Q})=SEC(\mathcal{P})$.
\end{definition}

\begin{figure}[!htbp]
\begin{center}
  \begin{minipage}[t]{0.3\linewidth}
    \centering
    \epsfig{file=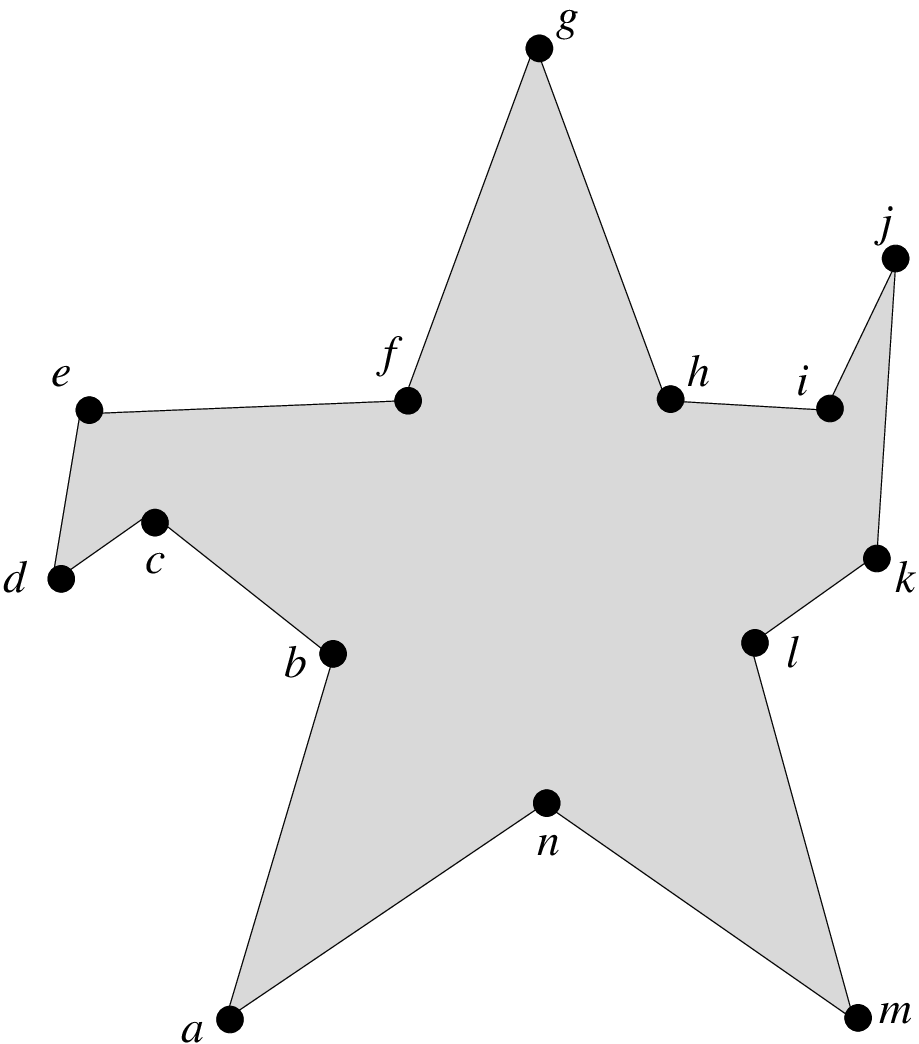, width=0.7\linewidth}\newline
    {\scriptsize ($a$) Positions~$a$ to~$n$ form the target pattern $\mathcal{P}$.}
  \end{minipage}
  \begin{minipage}[t]{0.3\linewidth}
    \centering
    \epsfig{file=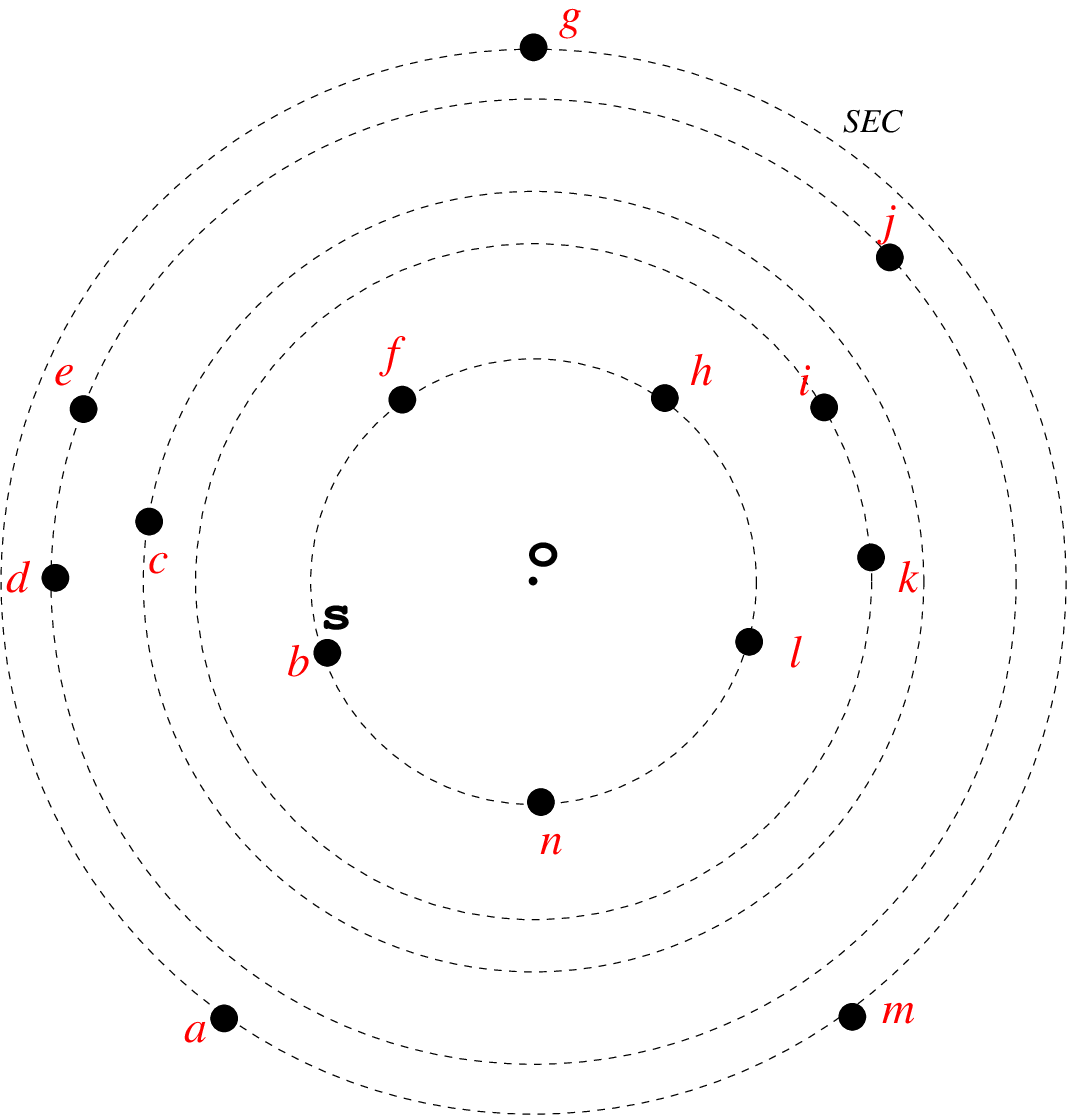, width=0.7\linewidth}\newline
    {\scriptsize ($b$) Position~$b$ is the first non-critical position $s$ in $\mathcal{P}$.}
  \end{minipage}
  \begin{minipage}[t]{0.3\linewidth}
    \centering
    \epsfig{file=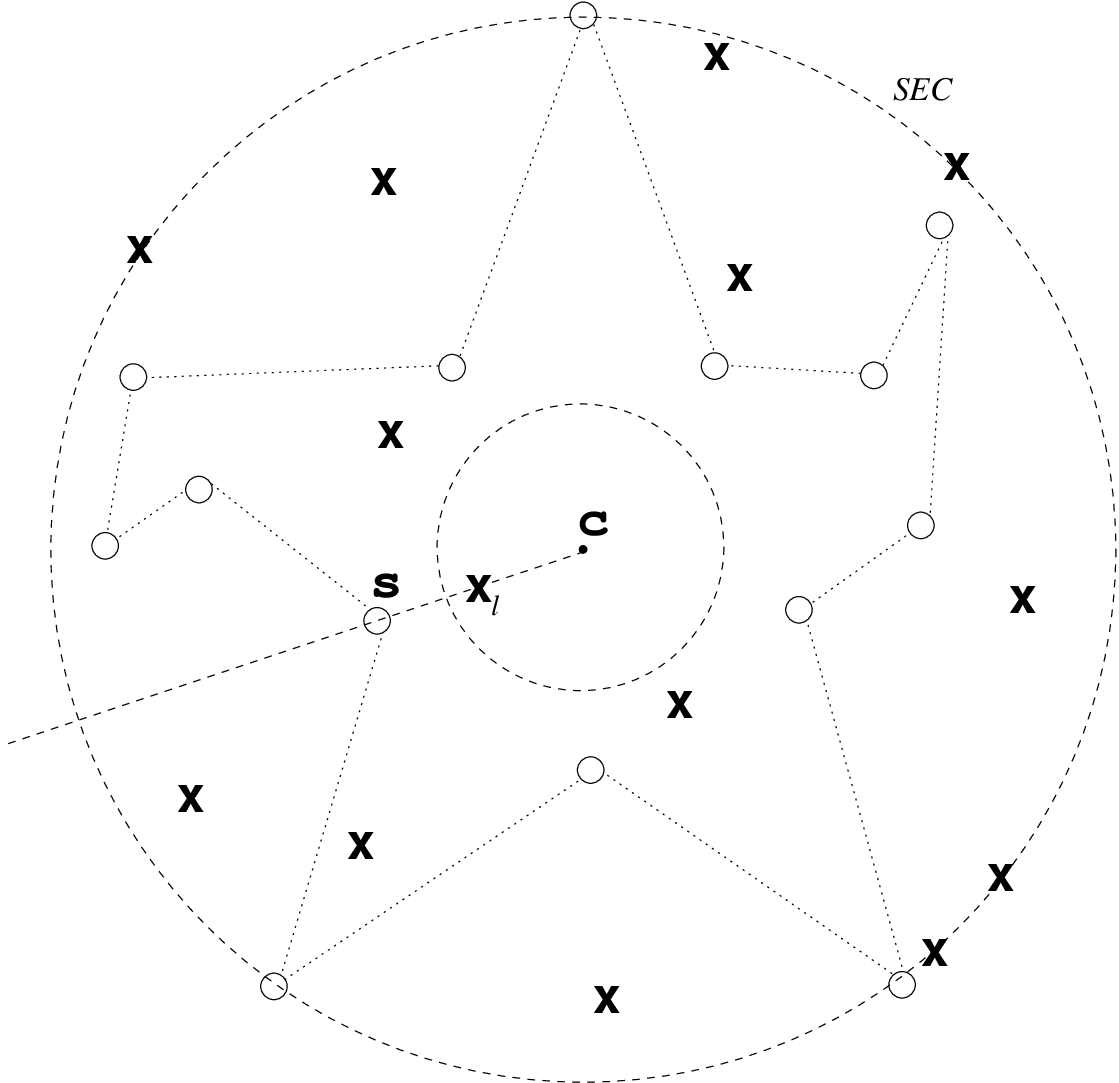, width=0.7\linewidth}\newline
    {\scriptsize ($c$) $\mathcal{P}$ mapped on the actual agreement configuration $\mathcal{Q}$ formed by the robots
(Robots are depicted as {\sf x}'s in the figure).}
  \end{minipage}
\end{center}
 \caption{An example showing a pattern $\mathcal{P}$ mapped on an agreement configuration
$\mathcal{Q}$---Definition~\ref{def:map}.} 
\label{fig:pattern1}
\end{figure}

An example showing the construction of Definition~\ref{def:map} is given in Figure~\ref{fig:pattern1}. 

\begin{definition}[$(k,\mathcal{P})$-partial pattern]
\label{ksub}
Let $\mathcal{Q}$ and $\mathcal{P}$ be respectively an agreement configuration formed by the robots in the plane and a target pattern. We say that:
\newline \noindent 1. $\mathcal{Q}$ is a $(0,\mathcal{P})$-partial pattern if the leader in $\mathcal{Q}$ is inside the smallest concentric enclosing circle of $Map(\mathcal{Q},\mathcal{P})$.
\newline \noindent 2. $\mathcal{Q}$ is a $(k,\mathcal{P})$-partial pattern with $1\leq k\leq Min(|\mathcal{SC}^{\mathcal{Q}}|,|\mathcal{SC}^{\mathcal{P}}|)$ if the three following properties holds: 
\newline {$\ ~~$} \hspace{1cm} a. $\mathcal{Q}$ is a $(0,\mathcal{P})$-partial pattern.
\newline {$\ ~~$} \hspace{1cm} b. $C^{Map(\mathcal{Q},\mathcal{P})}_{k}\cap Map(\mathcal{Q},\mathcal{P}) \subseteqq C^{\mathcal{Q}}_{k}\cap\mathcal{Q}$.
\newline {$\ ~~$} \hspace{1cm} c. $\bigcup_{i=1}^{k-1} C^{\mathcal{Q}}_i\cap\mathcal{Q}= \bigcup_{i=1}^{k-1} C^{Map(\mathcal{Q},\mathcal{P})}_i\cap Map(\mathcal{Q},\mathcal{P})$.
\end{definition}

In the sequel, we say that $\mathcal{Q}$ is a \emph{maximal} $(k,\mathcal{P})$-partial pattern if $\mathcal{Q}$ is a $(k,\mathcal{P})$-partial pattern and not a $(k+1,\mathcal{P})$-partial pattern.

\begin{definition}[Extra robots]
\label{def:extra}
Let $\mathcal{P}$ and $\mathcal{Q}$ be respectively a target pattern and a configuration formed by the robots in the plane  such that $\mathcal{Q}$ is a maximal $(k,\mathcal{P})$-partial pattern. We say that a robot $r$ is an extra robot if one of the two following properties holds:
\begin{enumerate}
\item $k=0$, $r$ is inside $SEC(\mathcal{Q})$, and $r$ is not the leader in $\mathcal{Q}$;

\item $k\geq1$ and 
\begin{enumerate}
\item either $r$ is inside the enclosing circle $C^{Map(\mathcal{Q},\mathcal{P})}_{k}$ and $r$ is not the leader in $\mathcal{Q}$;

\item or $r$ is on the circumference of $C^{Map(\mathcal{Q},\mathcal{P})}_{k}$ and $r$ does not occupy a position in  $C^{Map(\mathcal{Q},\mathcal{P})}_{k}\cap Map(\mathcal{Q},\mathcal{P})$.
\end{enumerate}
\end{enumerate} 
\end{definition}



\begin{algo}[htb!]
\begin{footnotesize}
\begin{tabbing}
  xxxxx \= xxxxx \= xxxxx \= xxxxx \= xxxxx \= xxxxx \= xxxxx \= xxxxx \= xxxxx \= \kill 
$\mathcal{Q}:=$ the configuration where the robots currently lies; \\
$\mathcal{P}:=$ the target pattern; /* $\mathcal{P}$ is the same for all the robots */ \\
$r_l:= Leader(\mathcal{Q})$; \\
$s:=$ the first non-critical position located on the smallest concentric enclosing \\
\> circle of $Map(\mathcal{Q},\mathcal{P})$;\\
\IF {the robots do not form any $(k,\mathcal{P})$-partial pattern}\\
\THEN  \> /*$r_l$ is not inside the smallest concentric enclosing circle of $Map(\mathcal{Q},\mathcal{P})$ */\\
        \> $p:=$ the middle of the segment $[c;s]$; \\
      \> \IF {I am $r_l$} \THEN $MoveTo(p,\rightarrow)$;\\
\ELSE \> /* the robots form a $(k,\mathcal{P})$-partial pattern */\\
\>\IF { the center of $SEC(\mathcal{Q})$ $\in Map(\mathcal{Q},\mathcal{P})$ } \\
\>\THEN $x:=$ the center of $SEC(\mathcal{Q})$;\\
\>\ELSE $x:=s$;\\
\> $Final\_Positions:=$ $Map(\mathcal{Q},\mathcal{P})\setminus \{x\};$\\
      \>  \IF {all the positions in $Final\_Positions$ are occupied}\\
      \>  \THEN \> \IF {I am $r_l$} \THEN $MoveTo(x,\rightarrow)$;\\
\> \ELSE \> $k:=$ the maximal $k$ for which $\mathcal{Q}$ is a $(k,\mathcal{P})$-partial pattern;\\
\> \> \IF {there is at least one extra robot not located on $C^{Map(\mathcal{Q},\mathcal{P})}_{k+1}$}\\
\> \> \THEN \> $r:=Nearest\_extra\_robot(C^{Map(\mathcal{Q},\mathcal{P})}_{k+1},\mathcal{Q},Map(\mathcal{Q},\mathcal{P}))$; \\
            \> \> \> $p:=Nearest\_free\_point(C^{Map(\mathcal{Q},\mathcal{P})}_{k+1},\mathcal{Q},r)$; \\
       \> \> \> \IF {I am $r$} \THEN $MoveTo(p,\rightarrow)$; \\
\> \> \ELSE $Arrange(C^{Map(\mathcal{Q},\mathcal{P})}_{k+1},Final\_Positions)$ 
\end{tabbing}
\end{footnotesize}
\caption{Procedure~$\AT$ for any robot $r_i$ in an agreement configuration \label{algo:AT}}
\end{algo}

The routine $Nearest\_extra\_robot(C^{Map(\mathcal{Q},\mathcal{P})}_{k+1},\mathcal{Q},Map(\mathcal{Q},\mathcal{P}))$ returns an extra robot $r$ such that $r$ is the closest extra robot to $C^{Map(\mathcal{Q},\mathcal{P})}_{k+1}$ which is not located on $C^{Map(\mathcal{Q},\mathcal{P})}_{k+1}$. If several candidates exists, then the extra robots inside $C^{Map(\mathcal{Q},\mathcal{P})}_{k+1}$ have priority.  Finally, if there is again several candidates then these latter ones are located on the same concentric circle $C$ centered at the center $c$ of $C^{Map(\mathcal{Q},\mathcal{P})}_{k+1}$ and the routine returns the extra robot, located on $C$, which is the closest in clockwise to the intersection between  $C$ and the half line $[c,r_l)$ (with $r_l$ the leader in $\mathcal{Q}$).

$Nearest\_free\_point(C^{Map(\mathcal{Q},\mathcal{P})}_{k+1},\mathcal{Q},r)$ returns the nearest position from $r$ \\ 
which is located on $C^{Map(\mathcal{Q},\mathcal{P})}_{k+1}$ and not occupied by any robot belonging to $\mathcal{Q}$. If there are two nearest positions then the routines returns the position which is the closest in clockwise to the intersection between  $C^{Map(\mathcal{Q},\mathcal{P})}_{k+1}$ and the half line $[c,r_l)$ (with $c$ the center of $C^{Map(\mathcal{Q},\mathcal{P})}_{k+1}$ and $r_l$ the leader). 

$MoveTo(p,C,\circlearrowright)$ allows a robot to move toward a position $p$ located on the circle $C$ by moving along the boundary of $C$ in clockwise. $MoveTo(p,C,\circlearrowleft)$ is similar but in counterclockwise.



$Arrange(C^{Map(\mathcal{Q},\mathcal{P})}_{k+1},Final\_Positions)$ allows all the robots on
$C^{Map(\mathcal{Q},\mathcal{P})}_{k+1}$ to occupy all the positions in $C^{Map(\mathcal{Q},\mathcal{P})}_{k+1}\cap
Final\_Positions$. 
The function is described by Algorithm~\ref{algo:AR} in which we use the following notions:

\begin{definition}[$arc(p,p',C,\circlearrowright)$]
\label{def:arc}
Given a circle $C$ and two points $p,p'$ located on it, $arc(p,p',C,\circlearrowright)$ is the arc of circle
$C$ from $p$ to $p'$ in the clockwise direction, $p$ being excluded ($p'$ being included). 
\end{definition}

\begin{definition}[$P$-$arc(p_i,p_{i+1},C,\circlearrowright)$]
Given a target pattern $\mathcal{P}$ and an agreement configuration $\mathcal{Q}$, we say that $arc(p_i,p_{i+1},C,\circlearrowright)$ is a $P$-$arc(p_i,p_{i+1},C,\circlearrowright)$ if and only if the three following properties holds:
\newline \noindent 1. $C$ is one of the concentric enclosing circle of $Map(\mathcal{Q},\mathcal{P})$
\newline \noindent 2. $p_i$ and $p_{i+1}$ belong to $Final\_Positions$
\newline \noindent 3. $p_{i+1}=adjacent(p_i,C,\circlearrowright)$
\end{definition}

\begin{remark}
\label{rem:DC}
 From Definition~\ref{def:arc}, we know that $p_i$ is not located on \\ $P$-$arc(p_i,p_{i+1},C,\circlearrowright)$.
\end{remark}

\begin{figure}[!htbp]
  \begin{center}
      \epsfig{file=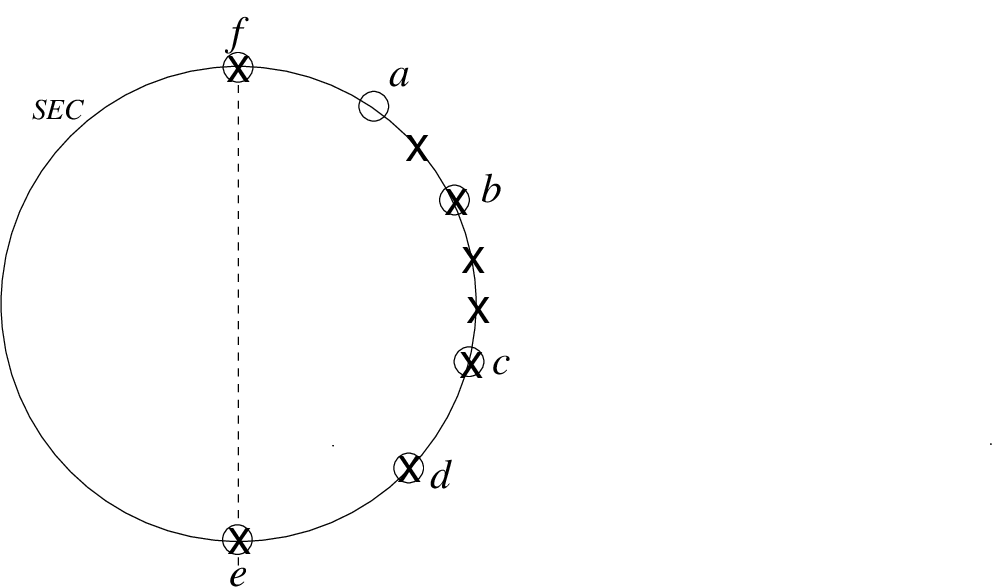, width=0.6\linewidth}
  \end{center}
  \caption{An example showing a Deadlock Chain and a Deadlock Breaker.} 
\label{fig:DC}
\end{figure}

In the remainder, we say that a $P$-arc is \emph{free} if there is no robot located on it.
In Figure~\ref{fig:DC}, the circles denote the positions to achieve.  The crosses depict the robots.  The $P$-arc
starting after $f$ ($f$ excluded) and finishing at $a$ is free. 

\begin{definition}[Deadlock Chain]
\label{def:DC}
A Deadlock Chain is a consecutive sequence of $P$-$arc$ starting from a free $P$-$arc$ $P_0$ and followed in
the counterclockwise direction by a $P$-$arc$ $P_1$ such that: 
\newline \noindent 1. $P_1$ is a $P$-$arc(p,p',C,\circlearrowright)$ such that $angle(p,c,p',\circlearrowright)=180^o$ and there is only one robot $r$ on it and $r$ is located at $p'$,
\newline \noindent 2. and $P_1$ is followed in counterclockwise by a consecutive sequence (possibly empty) of \\ $P$-$arc(p,p',C,\circlearrowright)$ such that there is only one robot $r$ on each of them and $r$ is located at $p'$,
\item and that consecutive sequence (possibly empty) is followed by a $P$-$arc(p,p',C,\circlearrowright)$ such that
there is at least two robots on it and one of them is located at $p'$. This $P$-$arc$ is called the last $P$-$arc$ of
the deadlock chain.
\end{definition}

In Figure~\ref{fig:DC}, the segment starting from Position $a$ ($a$ included) to Position $b$ ($b$ excluded)
forms a deadlock chain. 

\begin{definition}[Deadlock Breaker]
\label{def:DCB}
Let $P$-$arc(p,p',C,\circlearrowright)$ be the last $P$-$arc$ of a deadlock chain. The deadlock breaker is the robot located at $p'$. 
\end{definition}

In Figure~\ref{fig:DC}, the robot located at Position~$c$ is the deadlock breaker.

\begin{algo}[htb!]
\begin{footnotesize}
\begin{tabbing}
  xxxxx \= xxxxx \= xxxxx \= xxxxx \= xxxxx \= xxxxx \= xxxxx \= xxxxx \= xxxxx \= \kill 
/* I am $r_i$ */\\
 $p:=$ the closest position in $C^{Map(\mathcal{Q},\mathcal{P})}_{k+1}\cap Final\_Positions \setminus \{r_i\}$ to $r_i$ in clockwise; \\
\IF {$C^{Map(\mathcal{Q},\mathcal{P})}_{k+1}= SEC(\mathcal{Q})$} \\
\THEN \> \IF {there is no robot in $arc(r_i,p,C^{Map(\mathcal{Q},\mathcal{P})}_{k+1},\circlearrowright)$ or I am a
deadlock breaker }\\
\> \THEN \> \IF {I am a deadlock breaker} \\ 
\> \> \THEN $t:=$ the position s.t. $angle(r_i,c,t,\circlearrowright)=\frac{1}{2}angle(r_i,c,p,\circlearrowright)$;\\
\>\>\>$p:=t$;\\
\>\> \ENDIF \\
\> \> $r_{i-1}:= adjacent(r_i,SEC,\circlearrowleft)$; \\ 
    \>  \> $p':=$ the position such that $angle(r_{i-1},c,p',\circlearrowright)=180^o$;\\
     \> \> $p'':=$ the closest point to $r_i$ in clockwise in $\{p;p'\}$;\\  
     \> \> \IF {$r_i$ is not located at $p''$} \THEN $MoveTo(p'',SEC,\circlearrowright)$;\\
 \>\ENDIF \\
 \ELSE \> \IF {there is no robot in $arc(r_i,p,C^{Map(\mathcal{Q},\mathcal{P})}_{k+1},\circlearrowright)$}\\
\> \THEN $MoveTo(p,C^{Map(\mathcal{Q},\mathcal{P})}_{k+1},\circlearrowright)$;
\end{tabbing}
\end{footnotesize}
\caption{$Arrange(C^{Map(\mathcal{Q},\mathcal{P})}_{k+1},Final\_Positions)$ executed by robot $r_i$ on $C^{Map(\mathcal{Q},\mathcal{P})}_{k+1}$ \label{algo:AR}}
\end{algo}

\subsubsection{Correctness Proof of Algorithm~\ref{algo:main}}

We first show that by executing Algorithm~\ref{algo:main}, the smallest enclosing circle $SEC(\mathcal{Q})$ remains
invariant---Lemma~\ref{lem:inva}.

\begin{lemma}
\label{lem:inva}
According to Algorithm~\ref{algo:main}, the smallest enclosing circle $SEC(\mathcal{Q})$ remains invariant.
\end{lemma}

\begin{proof}
Assume by contradiction $SEC(\mathcal{Q})$ does not remain invariant. From Corollaries~\ref{cor:cri} and~\ref{theorem1} and Property~\ref{pro:sec}, we deduce  that can occurs if and only if:
\begin{itemize}
\item  Either a robot $r$ moves outside $SEC(\mathcal{Q})$. However, according to Algorithm~\ref{algo:main}, no robot moves outside $SEC(\mathcal{Q})$. That is a contradiction.
\item  Or an angle strictly greater than $180^o$ appears between two adjacent robots $r_{i-1}$ and $r_i$, i.e., $angle(r_{i-1},c,r_i,\circlearrowright)>180^o$ with $r_{i-1}= adjacent(r_i,SEC,\circlearrowleft)$. This subcase can occur if and only if 
\begin{itemize}
\item Either a critical robot leaves $SEC(\mathcal{Q})$. However, according to Algorithm~\ref{algo:LA} no critical
robot leaves $SEC(\mathcal{Q})$ ( only the first non-critical robot on $SEC$ in clockwise is sometimes allowed to
move). Furthermore, according to Algorithm~\ref{algo:AT} some robots are allowed to leave $SEC(\mathcal{Q})$  only if
these latter ones are extra robots and $\mathcal{Q}$ is a $(1,\mathcal{P})$-partial pattern. That implies some robots
are allowed to leave $SEC(\mathcal{Q})=C_1^{\mathcal{Q}}$ only if these latter ones do not occupy a position $\in
Map(\mathcal{Q},\mathcal{P})\cap SEC(\mathcal{Q})$ and  all the positions in $Map(\mathcal{Q},\mathcal{P})\cap
SEC(\mathcal{Q})$ are occupied by some robots. However, from Corollary~\ref{theorem1} we know that for all couple of
positions $r_i$ and $r_j$ on $Map(\mathcal{Q},\mathcal{P})\cap SEC(\mathcal{Q})$ such that
$r_j=adjacent(r_i,\circlearrowright)$, we have $angle(r_i,c,r_j,\circlearrowright)\leq180^o$. Consequently, when extra
robots leaves $SEC(\mathcal{Q})$,  $SEC(\mathcal{Q})$ is not changed. So, no critical robot leaves $SEC(\mathcal{Q})$. That is a contradiction. 
\item Or two adjacent robots $r_{i-1}$ and $r_i$, such that $r_{i-1}= adjacent(r_i,SEC,\circlearrowleft)$, move along $SEC(\mathcal{Q})$ so that  $angle(r_{i-1},c,r_i,\circlearrowright)>180^o$. That might occur only by applying Algorithm~\ref{algo:AR}. However, if $r_i$ is allowed to move, it can only move in clockwise towards a position $p$ such that $angle(r_{i-1},c,p,\circlearrowright)\leq180^o$. Furthermore, $r_{i-1}$ is never allowed to move in counterclockwise. So, $angle(r_{i-1},c,r_i,\circlearrowright)$ is always less than or equal to $180^0$. That is a contradiction.   
\end{itemize} 
\end{itemize}
\end{proof}

 From now on, we prove that if the robots form a leader configuration which is not a final pattern 
$\mathcal{P}$ and not an agreement configuration, they eventually form an agreement configuration---Lemma~\ref{lem:LA}.   

\begin{lemma}
\label{lem:LA}
If the robots form a leader configuration which is not a final pattern $\mathcal{P}$ and not an agreement configuration, they form an agreement configuration in a finite number of cycles.
\end{lemma}
\begin{proof}
If the robots form a leader configuration which is not a final pattern $\mathcal{P}$ and not an agreement configuration, then from Corollary~\ref{corollaire} we have two cases to consider: either $(1)$ the leader $r_l$ is at the center $c$ of $SEC$ or $(2)$ $r_l$ is not the unique robot closest to $c$. 

\begin{itemize}
\item {\bf Case~1}. $r_l$ is at the center of $SEC$. According to Procedure~$\LA$, $r_l$ moves away from $c$ towards a position which is closer to the center than the second robot closer to the center. Furthermore, from Lemma~\ref{lem:inva}, the center $c$ of $SEC$ remains invariant even if $r_l$ moves. So, $r_l$ remains the unique leader and, by fairness, we deduce that an agreement configuration is formed in a finite number of cycles.

\item {\bf Case~2}. $r_l$ is not the unique robot closest to $c$. In that case, we have two subcases to consider:

\begin{itemize}

\item {\bf Case~2.1}. $r_l$ is not a critical robot. In this subcase, $r_l$ moves towards a position which is located between $c$ and itself (except $c$ and itself). From Lemma~\ref{lem:inva}, the center $c$ of $SEC$ remains invariant even if $r_l$ moves. So, by fairness we know that an agreement configuration is formed in a finite number of cycles. 

\item {\bf Case~2.2}. $r_l$ is a critical robot. From Corollary~\ref{cor:cri}, $r_l$ is on the circumference of $SEC$. However, by assumption $r_l$ is also one of the robots closest to the center of $SEC$. So, we deduce that all the robots are on $SEC$. Hence, by  Lemma~\ref{lem:critical}, we deduce there is at least one non-critical robot on $SEC$ because there are at least four robots on it (recall that we assume the number of robot is greater than or equal to $4$). 

According to Procedure~$\LA$ the first non-critical robot $r_k$ starting from $r_l$ on $SEC$ in clockwise is allowed to move toward a position located between itself and $c$ (except $c$ and itself). From Lemma~\ref{lem:inva}, the center $c$ of $SEC$ remains invariant even if $r_k$ moves. So, by fairness $r_k$ becomes the unique robot closest to $c$ and it is not located at $c$.. So, the robots form an agreement configuration in a finite number of cycles.
\end{itemize} 
\end{itemize} 
\end{proof}

Starting from such a configuration, $Map(\mathcal{Q},\mathcal{P})$ remains invariant or the target pattern
$\mathcal{P}$ is formed---Lemma~\ref{lem:noref1} and Corollary~\ref{cor:mapinv}. Note that Corollary~\ref{cor:mapinv} assures that the two parts of Algorithm ~\ref{algo:main} ( Protocol~$\LA$ and Protocol~$\AT$) work in the asynchronous model $CORDA$ even if the unique robot closest to the center of $SEC$ is not still. 

\begin{lemma}
\label{lem:noref1}
Starting from an agreement configuration $\mathcal{Q}$, the robots remain in an equivalent agreement configuration or the target pattern $\mathcal{P}$ is formed in a finite number of cycles.
\end{lemma}
\begin{proof}
According to Lemma~\ref{lem:inva}, $SEC(\mathcal{Q})$ and its center $c$ remain invariant. Moreover, according to Algorithm~\ref{algo:main} and more precisely Algorithm~\ref{algo:AT} no robot is allowed to pass $r_l$. 

So, if $r_l$ is not allowed to move then, according to Definition~\ref{def:equiv} all the robots remain in an equivalent agreement configuration. 

If $r_l$ is allowed to move then, according to Algorithm~\ref{algo:AT} that can occur only in three cases: 

\begin{itemize}
\item {Case 1}. The robots do not form any $(k,\mathcal{P})$-partial pattern. In that case, $r_l$ moves in straight line towards the middle $p$ of the segment $[c,s]$ in order to get closer to the center $c$. However, from Definition~\ref{def:map}, we know that $s$ is on the half line $[c,r_l)$. So, during the motion of $r_l$, all the robots clearly remain in an equivalent agreement configuration.

\item {Case 2}. The center $c$ of $SEC(\mathcal{Q})$ is in $Map(\mathcal{Q},\mathcal{P})$ and all the positions in $Map(\mathcal{Q},\mathcal{P})$ are occupied except $c$. In that case, $r_l$ chooses to move towards $c$ in straight line (i.e., along $[c,r_l)$) in order to occupy the last free position in $Map(\mathcal{Q},\mathcal{P})$. Until $r_l$ has not reached $c$, the robots remain in an equivalent agreement configuration because $r_l$ is still on the same half line $[c,r_l)$ and it remains the unique robot closest to $c$. So by fairness, it reaches $c$ in a finite number of cycle and the pattern $\mathcal{P}$ is formed. 

\item {Case 3}. The center $c$ of $SEC(\mathcal{Q})$ is not in $Map(\mathcal{Q},\mathcal{P})$ and all the positions in $Map(\mathcal{Q},\mathcal{P})$ are occupied except the first non critical position $s$ located on the smallest concentric enclosing circle. In that case, $r_l$ chooses to move towards $s$ in order to occupy the last free position in $Map(\mathcal{Q},\mathcal{P})$. From Definition~\ref{def:map}, we know that $s$ is on the half line $[c,r_l)$ and thus, until $r_l$ has not reached $s$ the robots remain in an equivalent agreement configuration because $r_l$ is still on the same half line $[c,r_l)$ and it remains the unique robot closest to $c$. So by fairness, it reaches $s$ in a finite number of cycle and the pattern $\mathcal{P}$ is formed. 
\end{itemize}
\end{proof}

\begin{corollary}
\label{cor:mapinv}
 From an agreement configuration, $Map(\mathcal{Q},\mathcal{P})$ remains invariant or the target pattern $\mathcal{P}$ is formed.
\end{corollary}

It follows that from an agreement configuration which is not a 
$(k,\mathcal{P})$-partial pattern, the robots eventually form a $(0,\mathcal{P})$-partial
pattern---Lemma~\ref{lem:noref2}. 

\begin{lemma}
\label{lem:noref2}
 From an agreement configuration which is not a $(k,\mathcal{P})$-partial pattern, the robots form a $(0,\mathcal{P})$-partial pattern in a finite number of cycles.
\end{lemma}
\begin{proof}
 From Definition~\ref{ksub}, we know that if an agreement configuration is not a $(0,\mathcal{P})$-partial pattern then, the leader $r_l$ is not inside the smallest concentric enclosing circle of $Map(\mathcal{Q},\mathcal{P})$. From Corollary~\ref{cor:mapinv} and according to Algorithm~\ref{algo:AT}, $r_l$ is inside the smallest concentric enclosing circle of $Map(\mathcal{Q},\mathcal{P})$ in a finite number of cycles. 
\end{proof}

 From this point on, note that according to Algorithm~\ref{algo:AT}, $Final\_Positions$ is equal to all the positions in 
$Map(\mathcal{Q},\mathcal{P})$ except:
\newline \noindent 1. either the center $c$ of $SEC(\mathcal{Q})$ if $c\in Map(\mathcal{P},\mathcal{Q})$,
\newline \noindent 2. or the first non critical position located on the smallest concentric enclosing circle of $Map(\mathcal{Q},\mathcal{P})$ if $c\notin Map(\mathcal{P},\mathcal{Q})$

Now, we show by induction that, from a configuration being a maximal $(k,\mathcal{P})$-partial pattern, the robots
eventually form a $(k+1,\mathcal{P})$-partial pattern or the target pattern $\mathcal{P}$ is
formed---Lemmas~\ref{lem:inte} to~\ref{lem:best}.

\begin{lemma}
\label{lem:inte}
Let $\mathcal{P}$ be a target pattern and let $\mathcal{Q}$ be a configuration which is a maximal $(k,\mathcal{P})$-partial pattern such that 
$1\leq k<|\mathcal{SC}^\mathcal{P}|$. If all the extra robots are on $C_{k+1}^{Map(\mathcal{Q},\mathcal{P})}$ then, all the positions in $Final\_Positions\cap C_{k+1}^{Map(\mathcal{Q},\mathcal{P})}$ are occupied in a finite number of cycles.
\end{lemma}
\begin{proof}
If all the extra robots are on $C_{k+1}^{Map(\mathcal{Q},\mathcal{P})}$ and there exists at least one position in $Final\_Positions\cap C_{k+1}^{Map(\mathcal{Q},\mathcal{P})}$ which is not occupied then the robots apply the routine \\ $Arrange(C^{Map(\mathcal{Q},\mathcal{P})}_{k+1},Final\_Positions)$ (refer to Algorithm~\ref{algo:AR}). Remark that by applying this routine, no robot can collide with another robot since any robot can move only in clockwise and any move of a robot on $C_{k+1}^{Map(\mathcal{Q},\mathcal{P})}$ is only allowed in arc of circle containing no robot. Moreover, since $k\geq1$, $C_{k+1}^{Map(\mathcal{Q},\mathcal{P})}\ne SEC(\mathcal{Q})$ and thus, it is no need to prevent from creating an angle strictly greater than $180^o$ between two adjacent robots. In the remainder of this proof, we denote by $\alpha$ the number of extra robots located on $C_{k+1}^{Map(\mathcal{Q},\mathcal{P})}$, $\beta$ the number of $P$-arc on $C_{k+1}^{Map(\mathcal{Q},\mathcal{P})}$ and $\gamma$ the number of free $P$-arc on $C_{k+1}^{Map(\mathcal{Q},\mathcal{P})}$ .
According to Algorithm~\ref{algo:LA}, the acute reader noticed that the number $\alpha$ of extra robots is greater than or equal to the number $\beta$ of $P$-arc on $C_{k+1}^{Map(\mathcal{Q},\mathcal{P})}$.

We consider two cases.

\begin{itemize}
\item {\bf All the $P$-arcs are not free}. According to Algorithm~\ref{algo:AR}, each last robot on each $P$-$arc(p_i,p_{i+1},C,\circlearrowright)$ is allowed to move to $p_{i+1}$ if it is not yet at this position. At the end of these motions, all the positions in $Final\_Positions\cap C_{k+1}^{Map(\mathcal{Q},\mathcal{P})}$ are occupied and remains occupied.
\item {\bf At least one $P$-arc is free}. In that case we have $1\leq \gamma < \beta$. According to Algorithm~\ref{algo:AR}, if a robot moves from a $P$-arc to another one then $\gamma$ does not decrease because if robot $r$ chooses to move from a $P_1$-arc to a $P_2$-arc, that implies that $P_2$-arc is free. However, if $P_1$ becomes free when $r$ reaches $P_2$ then the number of free $P$-arc remains unchanged.

We now assume by contradiction that $\gamma$ never reaches the
value $\beta$.  So $\gamma$ eventually remains unchanged.  From this point on, no robot of any
$P$-$arc$ containing more than one robot will move towards a free $P$-arc.
Following the algorithm, that implies that every $P$-$arc$ $P$ with more than one robot is
followed in clockwise by a non free $P$-$arc$ infinitely often (at least each time the last robot of $P$
is awaked). Since the robots cannot move in counterclockwise, that also implies that every $P-arc$ with
more than one robot is always followed by a non free $P$-$arc$ $P'$.
If this second $P$-$arc$ $P'$ also contains more than one robot then it is also followed
by a non free $P$-$arc$.  However, if $P'$ contains only one robot then it is also followed by a
non free $P$-$arc$ $P''$, since on the contrary, the robot of $P'$ will eventually move to
$P''$ and the last robot of $P$ will eventually move to $P'$.  A contradiction.
So, step by step, it is clear that no $P$-$arc$ can be free and $\gamma=\beta$ which contradicts
our assumption.  So $\gamma$ will eventually reach the value $\beta$.

When $\gamma=\beta$ we retrieve the case where all the $P$-arcs are not free and the lemma holds.
\end{itemize}
\end{proof}

\begin{lemma}
\label{lem:secin}
Let $\mathcal{P}$ be a target pattern and let $\mathcal{Q}$ be a configuration which is a maximal $(0,\mathcal{P})$-partial pattern. If all the extra robots are on $SEC(Map(\mathcal{Q},\mathcal{P}))$ then, all the positions in $Final\_Positions\cap C_{k+1}^{Map(\mathcal{Q},\mathcal{P})}$ are occupied in a finite number of cycles.
\end{lemma}
\begin{proof}
If all the extra robots are on $SEC(Map(\mathcal{Q},\mathcal{P}))$ and there exists at least one position in
$Final\_Positions\cap SEC(Map(\mathcal{Q},\mathcal{P}))$ which is not occupied then the robots apply the routine \\
$Arrange(C^{Map(\mathcal{Q},\mathcal{P})}_{k+1},Final\_Positions)$ (refer to Algorithm~\ref{algo:AR}) for $k=1$.
Despite a more complicated code, the case $k=1$ can be seen as the case $k>1$ with an additionnal constraint on the
angles and a particular statement for a deadlock configuration removal. We show (refer to last item of this proof) that the deadlock removal generates a behavior that can finally be generated by Algorithm~\ref{algo:AR} for a concentric enclosing circle which is not $SEC(\mathcal{Q})$. So the aim of the proof is to show that Algorithm~\ref{algo:AR} has no deadlock.
In the rest of this proof we say that a point $p$ is a $P$-$point$ if $p \in Final\_Positions\cap SEC(Map(\mathcal{Q},\mathcal{P}))$.
So assume by contradiction that there exists a deadlock and we consider the two following cases:

\begin{enumerate}
  \item No $P-arc$ is free but there exists at least one $P$-$point$ which is not occupied by a robot.  Again, we distinguish two cases:
     \begin{enumerate} 
         \item At least one $P$-$point$ is occupied by a robot.  Let $p_i$ be one these $P$-$point$ such that
its successor in clockwise $p_{(i+1)}$ is free.  Clearly, \\ $angle(p_i, c, p_{(i+1)}, \circlearrowright) \leq
180$ (even if the first non critical $s$ does not belong to $Final\_Positions$ because, due to the fact $s$ is not critical, 
from Lemma\ref{lemma:sec} its absence cannot create an angle $>180$). So the last robot of the $p_{(i+1)}$
$P-arc$ can move to $ p_{(i+1)}$. A contradiction.
          \item No $P$-$point$ is occupied by a robot.  Since there are at least three robots on $SEC$, at
least one of them has a predecessor with an angle less than $180^o$.  So it can move.  A contradiction.
     \end{enumerate}
  \item There exists at least one free $P-arc$.  Let $i$ ($0 \leq i \leq \alpha - 1$) be an integer such that the $i th$ $P-arc$ is free and its predecessor (the $((i - 1) mod \alpha)^{th}$ $P-arc$) is not.  Let us call them $A$ and $A'$, respectively. We distinguish two cases:
     \begin{enumerate}
          \item There exists $(A',A)$ such that $A'$ contains at least two robots.  We call $r$ the last robot
of $A'$ and $r'$ the predecessor of $r$ on $A'$. In this case $r$ can move to $A$ (since $angle(r', c , r,
\circlearrowright) < angle(p_{(i - 2) mod \alpha}, c , p_{(i - 1) mod \alpha}, \circlearrowright) \leq 180^o$).  A contradiction.
          \item Every couple $(A',A)$ is such that $A'$ contains one robot only.  In that case there exists at
least a couple $(A', A)$ such that the predecessor $A''$ of $A'$ contains at least one robot since there are
at least as many robots as $P-points$ on $SEC$.  Because the deadlock assumption,  the robot $r$ on $A'$
cannot move to $A$ so $angle(r', c , r, \circlearrowright)=180^o$ where $r'$ is the last robot on $A''$.
		Again, we distinguish two cases:
                \begin{enumerate}
                       \item $r$ is not on $p_{(i - 1) mod \alpha}$.  In this case $r'$ also is not on $p_{(i - 2) mod \alpha}$ since $angle(p_{(i - 2) mod \alpha}, c , r, \circlearrowright) < 180$.  Since there is at least a third robot on $SEC$, this robot $r''$ is such that $angle(r'', c , r', \circlearrowright) < 180$ so $r'$ can move toward $p_{(i - 2) mod \alpha}$.  A contradiction.
                       \item $r$ is on $p_{(i - 1) mod \alpha}$.  In this case 
			$r'$ is also on $p_{(i - 2) mod \alpha}$ and \\
			$angle(p_{(i - 2) mod \alpha}, c , p_{(i - 1) mod \alpha}, \circlearrowright) = 180$.
			Since no robot can move, we can see that the configuration on $SEC$ is as follows: 
			$A$ is the first $P-arc$ of a chain starting from A in the conterclockwise such that any 
			$P-arc$ of this chain but $A$ contains a robot at its $P$-$point$, we call this chain $PC$.  
			The last $P-arc$ of $PC$ is followed by a free $P-arc$ ($A$ if there exists no other free $P-arc$). 
			Since no robot can move we can deduce that between this free $P-arc$ and $A$ (in the 
			conterclockwise) all the$P-arcs$ are free.  So all the robots are on $PC$ and there exists 
			at least one $P-arc$ of $PC$ which contains at least two robots.  
			Let $B$ be the first $P-arc$ of the chain (starting from $A$ in conterclockwise) such that 
			$B$ contains at least two robots.  Then the chain starting from $A$ and ending to $B$ is 
			a deadlock chain.  By definition, the robot on the $P$-$point$ of $B$ is a deadlock breaker
			and can move.  A contradiction.

			Now we just focus on the behavior of the successive deadlock breakers.  
			The aim of their behavior is to allow $r$ to move toward the next $P$-$point$.  
			It is easy to see that this part of the algorithm just reverses the order of the 
			deadlock breakers and $r$, but once ony of these robots has started to move 
			their behavior is the same as in the internal circle part (still with angle constraint).
			\end{enumerate}
		\end{enumerate}
\end{enumerate}
\end{proof}

\begin{lemma}
\label{lem:best}
Let $\mathcal{P}$ be a target pattern and let $\mathcal{Q}$ be a configuration which is a maximal $(k,\mathcal{P})$-partial pattern. The robots form a $(k+1,\mathcal{P})$-partial pattern or the target pattern is formed, in a finite number of cycles.
\end{lemma}
\begin{proof}
We have to consider three cases.
\begin{itemize}
\item {\bf $k=|\mathcal{SC}^{Map(\mathcal{Q},\mathcal{P})}|$.} In that case, $C_{k+1}^{Map(\mathcal{Q},\mathcal{P})}$ does not exist and \\
 $\bigcup_{i=1}^{|\mathcal{SC}^{Map(\mathcal{Q},\mathcal{P})}|} C^{\mathcal{Q}}_i\cap\mathcal{Q}= \bigcup_{i=1}^{|\mathcal{SC}^{Map(\mathcal{Q},\mathcal{P})}|} C^{Map(\mathcal{Q},\mathcal{P})}_i\cap Map(\mathcal{Q},\mathcal{P})$. That implies that it remains only one position $p$ to occupy and $p$ is inevitably at the center of $SEC(\mathcal{Q})$ (otherwise $C_{k+1}^{Map(\mathcal{Q},\mathcal{P})}$ would exist). According to Algorithm~\ref{algo:LA}, leader $r_l$ moves toward $c$. From Corollary~\ref{cor:mapinv} and by fairness, we deduce that the target pattern is formed in a finite number of cycles.

\item {\bf $k=|\mathcal{SC}^{Map(\mathcal{Q},\mathcal{P})}|-1$.} In that case, we distinguish two subcases:

\begin{enumerate}
\item The center $c$ of $SEC(\mathcal{Q})$ is in $Map(\mathcal{Q},\mathcal{P})$. In that subcase, all the positions in $C_{k+1}^{Map(\mathcal{Q},\mathcal{P})}\cap Map(\mathcal{Q},\mathcal{P}) $ must be occupy by all the extra robots even the first non critical position. According to Algorithm~\ref{algo:LA}, the extra robots move to the boundary of 
$C_{k+1}^{Map(\mathcal{Q},\mathcal{P})}$ by using subroutines
$Nearest\_extra\_robot(C^{Map(\mathcal{Q},\mathcal{P})}_{k+1},\mathcal{Q},Map(\mathcal{Q},\mathcal{P}))$ and \\
$Nearest\_free\_point(C^{Map(\mathcal{Q},\mathcal{P})}_{k+1},\mathcal{Q},r)$. These subroutines assure us that the
extra robots moves one by one toward a position on $C_{k+1}^{Map(\mathcal{Q},\mathcal{P})}$ which is not occupied by
any robot. Of course, if we are lucky, a $(k+1,\mathcal{P})$-partial pattern is formed during this step. Otherwise, the robots apply Algorithm~\ref{algo:AR} and, from Lemma~\ref{lem:inte} and \ref{lem:secin}, the $(k+1,\mathcal{P})$-partial pattern is formed in a finite number of cycles.

\item The center $c$ of $SEC(\mathcal{Q})$ is not in $Map(\mathcal{Q},\mathcal{P})$. In that subcase, all the positions in $C_{k+1}^{Map(\mathcal{Q},\mathcal{P})}\cap Map(\mathcal{Q},\mathcal{P}) $ must be occupy by all the extra robots except the first non critical position on $C_{k+1}^{Map(\mathcal{Q},\mathcal{P})}$ which is booked for the leader. According to Algorithm~\ref{algo:LA}, the extra robots move to the boundary of 
$C_{k+1}^{Map(\mathcal{Q},\mathcal{P})}$ by using subroutines \\
$Nearest\_extra\_robot(C^{Map(\mathcal{Q},\mathcal{P})}_{k+1},\mathcal{Q},Map(\mathcal{Q},\mathcal{P}))$ and \\
$Nearest\_free\_point(C^{Map(\mathcal{Q},\mathcal{P})}_{k+1},\mathcal{Q},r)$. During this step, if we are lucky, all
the positions in $C_{k+1}^{Map(\mathcal{Q},\mathcal{P})}\cap Map(\mathcal{Q},\mathcal{P}) $ are occupy by all the extra robots except the first non critical position on $C_{k+1}^{Map(\mathcal{Q},\mathcal{P})}$. Otherwise, the robtots apply Algorithm~\ref{algo:AR} and, from Lemmas~\ref{lem:inte} and \ref{lem:secin} all the positions in $C_{k+1}^{Map(\mathcal{Q},\mathcal{P})}\cap Map(\mathcal{Q},\mathcal{P})$ are eventually occupied except the first non critical position. From this point now, according to Algorithm~\ref{algo:LA} leader $r_l$ moves towards the first non critical position in $C_{k+1}^{Map(\mathcal{Q},\mathcal{P})}\cap Map(\mathcal{Q},\mathcal{P})$ and from Corollary~\ref{cor:mapinv} and fairness we deduce that the target pattern is formed in a finite number of cycles.
\end{enumerate}
\item {\bf $k<|\mathcal{SC}^{Map(\mathcal{Q},\mathcal{P})}|-1$.} In that subcase, all the positions in $C_{k+1}^{Map(\mathcal{Q},\mathcal{P})}\cap Map(\mathcal{Q},\mathcal{P}) $ must be occupy by all the extra robots. According to Algorithm~\ref{algo:LA}, the extra robots move to the boundary of 
$C_{k+1}^{Map(\mathcal{Q},\mathcal{P})}$ by using subroutines
$Nearest\_extra\_robot(C^{Map(\mathcal{Q},\mathcal{P})}_{k+1},\mathcal{Q},Map(\mathcal{Q},\mathcal{P}))$ and \\
$Nearest\_free\_point(C^{Map(\mathcal{Q},\mathcal{P})}_{k+1},\mathcal{Q},r)$. If we are lucky, a
$(k+1,\mathcal{P})$-partial pattern is formed during this step. Otherwise, the robots apply Algorithm~\ref{algo:AR} and, from Lemma~\ref{lem:inte} and \ref{lem:secin}, the $(k+1,\mathcal{P})$-partial pattern is formed in a finite number of cycles.
\end{itemize}
\end{proof}

 From Lemma~\ref{lem:best} and by induction we deduce the following theorem:
\begin{theorem}
Starting from a leader configuration, Algorithm~\ref{algo:main} allows to solve $APFP$ in $CORDA$ among a group of $n\geq 4$ robots having chirality and devoid of any kind of sense direction.
\end{theorem}

\begin{remark}
\label{rem:the}
Notice that our solution with chirality would not always guarantee the invariance of $SEC$ if $n=3$ and all the robots are placed on it. Indeed, in this
particular case, if there does not exists two robots that are on the same diameter it would be impossible to remove
one of the three without creating an angle greater than $180^o$ (which is not the case when $n\geq4$).  This is why
the given solution only works if we have four robots or more.
\end{remark}

\section{Equivalence without chirality for $n\geq5$}
\label{sec:sans}

What about the case without chirality ?

In such a context the result of Flocchini et al, summarized by Theorem~\ref{theo:implique} (i.e. the fact that it is possible to elect a leader for $n\geq3$ robots if it is possible to form any pattern for $n\geq3$), still holds. Therefore, in the same way as the chirality case, it remains to design an algorithm allowing to form an arbitrary pattern starting from a leader configuration. To reach such a design, it is worth noting that the major point we have to face lies in ensuring that, at some point, the robots have to agree on a common coordinate system. So, in the rest of this section, we only focus on how to achieve that kind of agreement. Combinated with the solution with chirality, the general scheme of the algorithm can be easily deduced.

Previously, when the robots shared the same chirality, the global coordinate system was implicitely determined by several ingredients namely: the center $c$ of $SEC$, the common chirality, and the unique position $r_{l_1}$ closest to $c$ which was occupied by the robot designated as leader.

For the case where the robots are devoided of a common handedness, it is possible to involve the same ingredients but by filling the lack of chirality by using the position of a second robot $r_{l_2}$ for which we will arrange that it is the unique robot closest to $c$ just after $r_{l_1}$ and such that $r_{l_1}$, $r_{l_2}$ and $c$ are not aligned. The chirality is then given by the orientation of the convex angle centered at $c$ between $r_{l_1}$ and $r_{l_2}$.

Once such a thihedron involving $r_{l_1}$, $r_{l_2}$ and $c$ is obtained, we can apply the same solution as the case with chirality by taking heed of some technicalities adressed in Subsection~\ref{sub:oth}.

However, to put in place this kind of thrihedron, we have to take care to overcome the two following problems which then arise. 

The first problem, referred to as Problem~1 below, consists in having a unique robot closest to $c$. This feature cannot be achieved in the same manner as when the robots share the same handedness especially because the set of leader configurations, in a context in which the robots have chirality, differs from the one when they have not. The second problem, referred to as Problem~2 below, consists in making the robots agree on a common handedness (By contrast with the previous case in which a common handedness is reached \emph{de facto}). 

The two next subsections are dedicated to describing solutions about both these problems.

\subsection{Solving Problem~1}
With chirality, reaching a configuration where only one robot is closest to $c$ is relatively an easy thing as leader configurations, in such a context, initially allow to distinguish a robot among those that are closest to $c$ (cf Corollary~\ref{corollaire}). However, without chirality, this is not always the case: There are some leader configurations in which all the robots that are potentially distinguishable are not included among those that are closest to $c$. In fact, with no chirality we have the following corollary.

\begin{corollary}\cite{DLPV12}
Let $\mathcal{Q}$ be a leader configuration of a set of robots devoided of a common handedness.
\begin{itemize}
\item If $\mathcal{Q}$ has no symmetry axis then we can distinguish a unique robot among those that are closest to $c$ (Case~1).
\item Otherwise, $Q$ has exactly one symmetry axis $S$ with at least one robot on it and, without any further move, all the distinguishable robots are on $S$. In particular, we can distinguish a robot which is among those that are closest to $c$ on $S$ (Case~2).  
\end{itemize}
\end{corollary}

From the above corollary, we then have two cases to consider, i.e. Case~1 and Case~2.
In the first case, if the leader $r_l$ is the unique robot closest to $c$ then the problem is over. Otherwise, either $r_l$  is not critical, in which case it is sufficient for it to slightly move towards $c$, or it is critical. In the latter case, we can anyway remark that since $r_l$ is not located on a symmetry axis, the robots can agree on some same total order over the set of all the positions in which $r_l$ is viewed as the first robot. Hence, as for the chirality case, the next non critical robot, which is immediately after $r_l$ among those that are closest to $c$, slightly moves towards $c$ and becomes the new leader as soon as it starts its motion since all the robots are on $SEC$ (indeed since $r_l$ is critical that implies $r_l$ is on $SEC$. Moreover, no robot can be closer to $c$ than $r_l$. So we know that all the robots are on the circumference of $SEC$). To do this, as stated in Lemma~\ref{lem:critical}, we have to assume there is at least $4$ robots in the team (however, as we shall see below in the resolution of Problem~2, we will even need to have at least $5$ robots).

In the second case, $r_l$ is a robot which is among those that are closest to $c$ on the single symmetry axis $S$. 

If $r_l$ is not critical or it is the unique robot closest to $c$ among those located on $S$ then it is enough that $r_l$ moves towards $c$ (the center of $SEC$ is also located on $S$) in order to become the unique robot closest to $c$. During its motion, $r_l$ cannot lose its leadership because

\begin{itemize}
\item if there no symmetry axis then $r_l$ is not critical and immediately becomes the unique robot closest to $c$ as soon as it moves.

\item And if there is exactly one symmetry axis $S$, then
\begin{itemize}
\item either $r_l$ is the unique robot on $S$, in which case it always remains the unique robot on $S$ and thus keeps its leadership 
\item or there is several robot on $S$. However since $r_l$ is the unique robot closest to $c$ among those that are on $S$, it is not critical. Hence, while moving to $c$, it remains the unique robot closest to $c$ among those that are on $S$ and also keeps its leadership.
\end{itemize} 
\end{itemize}

Otherwise, $r_l$ is critical and there is another robot, denoted $r_{l2}$ on $S$ which is located at the same distance from $c$: Actually, $r_l$ and $r_{l2}$ are diametrically opposite on the circumference of $SEC$. There is two subcases to consider.

If $r_{l2}$ is not critical, $r_{l2}$ moves towards the center of $SEC$ (as soon as it moves $r_{l2}$ becomes the new leader because it immediately becomes the closest robot to $c$ among those that are on $S$). 

If $r_{l2}$ is critical then $r_l$ moves to the boundary of the smallest enclosing circle $SEC_2$ of the set of all the robots, except the $r_l$, and passing through $r_{l2}$. In this way, once $r_l$ is on $SEC_2$ we retrieve a case where $r_l$ or $r_{l2}$ is not critical since $SEC_2$ now contains at least $4$ robots (cf Lemma~\ref{lem:critical}). Concerning this subcase, we would like to emphasize that $r_l$ remains leader while it moves towards the boundary of $SEC_2$. Indeed according to \cite{DLPV12}, if $r_l$ is leader while $r_{l2}$ is also on the unique symmetry axis and at the same distance to $c$, this is due to the fact that the angle $\alpha_1$ centered at $c$ between the two nearest radii from $r_l$, both of them passing through at least one robot different to $r_l$, is smaller than or equal to the angle $\alpha_2$ centered at $c$ between the two nearest radii from $r_{l2}$, both of them passing through at least one robot different to $r_{l2}$. However during the motion of $r_l$ to the circumference of $SEC_2$, $\alpha_1$ decreases while $\alpha_2$ increases. And since $r_l$ and $r_{l2}$ remain on the circumference of the current $SEC$ as well as they are equidistant of the center of the current $SEC$, $r_l$ keeps its leadership until it reaches the boundary of $SEC_2$.

\subsection{Solving Problem~2}

Once the first problem discussed above is solved, solving Problem~2 is straightforward. The main idea consists in ensuring that a robot $r_{l_2}$ is the unique second robot closest to $c$ and that $r_{l_2}$, $c$ and the unique robot $r_{l_1}$ closest to $c$ are not aligned. Indeed, in such a case the chirality can be given by the orientation of the convex angle centered at $c$ from $r_{l_1}$ to $r_{l_2}$ (or the converse according to the third technicality below). To achieve that, we have two cases to consider. 

If the configuration has no symmetry axis, thus by applying a similar strategy as before (to have a unique robot closest to $c$) we can obtain the desired configuration. Note that, like $r_{l_1}$, $r_{l_2}$ must be a non critical robot. By using a similar argument as that of Lemma~\ref{lem:critical} and from the fact that $r_{l_1}$ is not on $SEC$ when solving Problem~2, this condition can be fulfilled only when $n\geq5$, which corresponds to the lower bound of our solution without chirality.  

Otherwise, $r_{l_1}$ can break the symmetry by slightly moving along the circle centered at $c$ on which it is located, which brings us to the case without symmetry axis.

\subsection{Some other technicalities}
\label{sub:oth}
In closing, we would draw attention on three technicalities. 
First, as for the chirality case, it is appropriate to have $r_{l_1}$ and $r_{l_2}$ inside the minimal smallest concentric enclosing circle between that of $Map(\mathcal{Q},\mathcal{P})$ and $\mathcal{Q}$, in order to maintain invariant the coordinate system while the other robots move to their final positions. To do this, it is enough to place $r_{l_1}$ (resp. $r_{l_2}$) such that the distance between $c$ and $r_{l_1}$ is equal to half (resp. three quarter) the radius of the minimal smallest concentric enclosing circle between that of $Map(\mathcal{Q},\mathcal{P})$ and $\mathcal{Q}$. The second technicality is that we have to make sure that the coordinate system is stable when the robots, other than $r_{l_1}$ and $r_{l_2}$, start moving to the final positions of $Map(\mathcal{Q},\mathcal{P})$. To achieve that, it is sufficient to use a predicat which prevents the other robots from moving until $r_{l_1}$ and $r_{l_2}$ occupy their respective positions adressed above in the first technicality. Finally, the last technicality is about the way to occupy the final positions. Once the robots all share a common coordinate system via $r_{l_1}$ and $r_{l_2}$, the robots not defining this system move as for the chirality case. As for the chirality case again, $r_{l_1}$ has a reserved position $p_1$ located on the smallest concentric enclosing circle of $Map(\mathcal{Q},\mathcal{P})$ that it reachs when $\mathcal{Q}\setminus\{r_{l_1}\}\cup\{p_1\}$ corresponds to $\mathcal{P}$ for one of both orientations of $SEC(\mathcal{Q})$. Concerning $r_{l_2}$, its reserved position $p_2$ is a non critical position located either on the smallest concentric circle of $Map(\mathcal{Q},\mathcal{P})$ if there are at least two positions to occupy on it or located on the second smallest concentric circle of $Map(\mathcal{Q},\mathcal{P})$ otherwise. As soon as $\mathcal{Q}\setminus\{r_{l_1},r_{l_2}\}$ corresponds to $SEC(\mathcal{Q})\setminus\{p_1,p_2\}$, $r_{l_2}$ moves to $p_2$. Of course, we will have taken care that the motion of $r_{l_2}$ to $p_2$ is done without changing the common chirality. This constrainst can be respected by choosing the appropriate chirality when defining the common system: The chirality is given by the orientation of the convex angle centered at $c$ either from $r_{l_1}$ to $r_{l_2}$ or from $r_{l_2}$ to $r_{l_1}$ depending on the constraint (from $r_{l_1}$ to $r_{l_2}$ if the two solutions are equivalent).

Hence, we have the following theorem.

\begin{theorem}
\label{theo:result42}
In $CORDA$, assuming a group of $n \geq 5$ robots having no chirality and devoid of 
any kind of sense of direction, $LEP$ is solvable if and only if $APFP$ is solvable.
\end{theorem}

\section{Conclusion}
\label{sec:concl}

We studied the relationship between $APFP$ and $LEP$ among robots in $CORDA$.
We provided solutions allowing to form an arbitrary pattern starting from any geometric configuration wherein the leader election is possible. More precisely, our solutions work for four or more robots with chirality and for at least five robots without chirality.
Combined with the result in \cite{FPSW08}, we deduce that $APFP$ and $LEP$ are equivalent, \ie 
it is possible to solve $APFP$ for $n \geq4$ with chirality (resp. $n \geq5$ without chirality) if and only if $LEP$ is solvable too.
The possible equivalence for $n=3$ with chirality or for $n=3$ and $n=4$ without chirality remains an open problem. 
In a future work, we would like to investigate these cases in order to obtain a fully complete result.

\bibliographystyle{plain}
\bibliography{ngon}
\end{document}